\documentclass[superscriptaddress,amsmath,amssymb,aps,notitlepage,twocolumn,floatfix]{revtex4-1}
\usepackage{cmbright}
\DeclareFontShape{OT1}{cmss}{m}{it}{<->ssub*cmss/m/sl}{}

\newcommand{\uproman}[1]{\uppercase\expandafter{\romannumeral#1}}

\DeclareFontFamily{OT1}{cmbr}{\hyphenchar\font45 }
\DeclareFontShape{OT1}{cmbr}{m}{n}{%
	<-9>cmbr8
	<9-10>cmbr9
	<10-17>cmbr10
	<17->cmbr17
}{}
\DeclareFontShape{OT1}{cmbr}{m}{sl}{%
	<-9>cmbrsl8
	<9-10>cmbrsl9
	<10-17>cmbrsl10
	<17->cmbrsl17
}{}
\DeclareFontShape{OT1}{cmbr}{m}{it}{%
	<->ssub*cmbr/m/sl
}{}
\DeclareFontShape{OT1}{cmbr}{b}{n}{%
	<->ssub*cmbr/bx/n
}{}
\DeclareFontShape{OT1}{cmbr}{bx}{n}{%
	<->cmbrbx10
}{}

\usepackage[noindentafter]{titlesec}
\usepackage{titlesec}
\titleformat{name=\section}
{\normalfont\large\bfseries\MakeUppercase}{\MakeUppercase{\thesection}}{0pt}{}
\titleformat{name=\subsection}
{\normalfont\bfseries}{\thesection}{0pt}{}
\titlespacing{\section}{0cm}{0.7cm}{0.01cm}
\titlespacing{\subsection}{0cm}{0.45cm}{0cm}

\usepackage[utf8]{inputenc}
\usepackage{listings}
\usepackage{footmisc}
\usepackage{enumerate}
\usepackage{latexsym}
\usepackage{braket}
\usepackage{bm} % bold math-mode
\usepackage[version=4]{mhchem}
\usepackage{graphicx}
\usepackage{subfigure}
\usepackage[colorlinks=True,linkcolor=red,citecolor=blue,urlcolor=blue]{hyperref}
\usepackage[capitalise]{cleveref}
\usepackage{appendix}
\usepackage{blkarray}
\usepackage{array}
\usepackage[dvipsnames]{xcolor}
\usepackage[normalem]{ulem}
\usepackage{wasysym}
\usepackage{multirow}
\usepackage{overpic}

\usepackage{caption}
\usepackage{lipsum}

\usepackage{xspace}
\usepackage{comment}

\usepackage{tabularx}
\usepackage{array}   
\newcolumntype{L}{>{$}l<{$}} 
\newcolumntype{R}{>{$}r<{$}} 
\newcolumntype{C}{>{$}c<{$}} 

\usepackage{float}

\usepackage[capitalise]{cleveref}
\usepackage{placeins}

\usepackage[normalem]{ulem}

\date{\today}

\usepackage{comment}

\begin{document}
	\title{Revealing the microscopic origin of the magnetization plateau in  Na\textsubscript{3}Ni\textsubscript{2}BiO\textsubscript{6}}
	
	\author{Amanda A. Konieczna}
	\email{konieczna@itp.uni-frankfurt.de}
	\affiliation{Institut f\"ur Theoretische Physik, Goethe-Universit\"at, 60438 Frankfurt am Main, Germany}
	\author{P. Peter Stavropoulos}
	\email{panagiotis@itp.uni-frankfurt.de}
	\affiliation{Institut f\"ur Theoretische Physik, Goethe-Universit\"at, 60438 Frankfurt am Main, Germany}	
	\author{Roser Valent\'i}
	\email{valenti@itp.uni-frankfurt.de}
	\affiliation{Institut f\"ur Theoretische Physik, Goethe-Universit\"at, 60438 Frankfurt am Main, Germany}
	
	\date{\today}
	
	\begin{abstract}
		Recent experimental studies of the spin-1 honeycomb antiferromagnet Na$_3$Ni$_2$BiO$_6$ have revealed a pronounced one-third magnetization plateau under applied magnetic fields, highlighting the presence of strong magnetic frustration and anisotropy in this material. Such behavior has been attributed to substantial bond-dependent Kitaev interactions in combination with single-ion anisotropy, placing Na$_3$Ni$_2$BiO$_6$ among honeycomb compounds of interest for unconventional magnetic phases. Motivated by these observations, we present a first-principles–based analysis of the magnetic interactions in Na$_3$Ni$_2$BiO$_6$. By combining density-functional calculations with microscopic modelling, we extract the relevant exchange parameters and construct an effective spin model that quantitatively reproduces both the elastic neutron-scattering spectra and the magnetization curve. The model captures the experimentally observed zero-field zigzag magnetic order, and proposes a \textit{double-zigzag} state at intermediate magnetic fields, realizing the one-third magnetization plateau in a simpler way than suggested in previous works. Crucially, we show that the one-third magnetization plateau does not require Kitaev interactions; instead, it arises from the interplay of strong out-of-plane single-ion anisotropy and competing ferromagnetic nearest-neighbor ($J_1$) and antiferromagnetic third-neighbor ($J_3$) Heisenberg couplings. These results establish a consistent microscopic description of Na$_3$Ni$_2$BiO$_6$ and clarify the origin of its field-induced plateau phase.
	\end{abstract}

	\maketitle
	
	\section{INTRODUCTION}
	
	\vspace{0.2cm}

    For decades, frustration in magnetic systems has served as a central platform for exploring intriguing unconventional properties~\cite{balents2010spin,savary2016quantum}. It leads to highly degenerate ground states, suppressing conventional magnetic order and potentially giving rise to strongly entangled, exotic states. Possible emergent phases include quantum-spin-liquids~\cite{balents2010spin,savary2016quantum,rousochatzakis2024beyond,broholm2020quantum,wen2019experimental}, non-collinear magnetic structures~\cite{starykh2015unusual}, and nematic phases~\cite{fernandes2014drives,glasbrenner2015effect,scherer2017interplay}. Magnetically frustrated systems also show magnetic-field induced phases with characteristic magnetization plateaus~\cite{fortune2009cascade,zhou2012successive,susuki2013magnetization,kamiya2018nature,zhitomirsky2002field,damle2006spin,nishimoto2013controlling,yoshida2022frustrated,heidrich2006frustrated,jeschke2011multistep,heinze2021magnetization}, as observed in spin-$1/2$ systems on triangular~\cite{fortune2009cascade,zhou2012successive,susuki2013magnetization,kamiya2018nature} and kagome lattices~\cite{zhitomirsky2002field,damle2006spin,nishimoto2013controlling,yoshida2022frustrated}, where frustration originates from the lattice geometry. 
    
    Alternatively, even though the honeycomb lattice itself is not geometrically frustrated, frustration can emerge from bond-dependent magnetic interactions, as exemplified by Kitaev material candidates~\cite{singh2010antiferromagnetic,plumb2014alpha,winter2017models,trebst2022kitaev,rousochatzakis2024beyond} that realize the Kitaev model on the honeycomb lattice~\cite{kitaev2006anyons}. In these systems, nearest-neighbor (NN) bond-dependent Ising type spin-spin interactions are the source of magnetic frustration. The ground state of the pure Kitaev model is a quantum spin liquid, featuring emergent $\mathbb{Z}_2$ gauge fields coupled to itinerant Majorana fermions~\cite{kitaev2006anyons}. Real materials with Kitaev interactions often also host additional bond-dependent spin couplings~\cite{rau2014generic,winter2016challenges,yadav2016kitaev,kim2016crystal,rethinking}, which can stabilize a rich variety of phases, particularly under applied magnetic fields~\cite{winter2018probing,hickey2019emergence,kaib2019kitaev,ponomaryov2020nature,sahasrabudhe2020high,sarkis2026intermediate,park2021hinge,feng2026magnetic}.
 	Recent experiments in the spin-1 honeycomb compound Na$_3$Ni$_2$BiO$_6$ (NNBO)~\cite{shangguan2023onethirdmag, shi2024low} have reported the appearance of a one-third magnetization plateau at intermediate magnetic fields, which has been attributed to an unusually strong Kitaev exchange coupling, as is well established in honeycomb compounds such as $\alpha$-RuCl$_3$ or $A_2$IrO$_3$. However, the microscopic origin of the magnetic exchanges in NNBO, unlike these systems, is unclear. 
    
	In this work, we investigate the origin of the exchange interactions in NNBO by performing full relativistic density functional theory (DFT) simulations combined with exact diagonalization of finite clusters and projection methods (projED)~\cite{riedl2019ab,kaib2022electronic,konieczna2025understanding}. We find, in contrast to previous literature, that the magnetic couplings in NNBO are dominated by 1\textsuperscript{st} NN and 3\textsuperscript{rd} NN Heisenberg interactions, with only a small contribution from Kitaev interactions, as well as a considerable single-ion anisotropy (SIA). Within our model, the occurrence of the field-induced one-third magnetization plateau is explained, and experimental elastic neutron spectroscopy (ENS) measurements are fully replicated.  We argue that the interplay between ferro- and antiferromagnetic Heisenberg couplings on 1\textsuperscript{st} and 3\textsuperscript{rd} NN is sufficient to explain the observations of the magnetization plateau, without invoking large bond-dependent couplings.
	\begin{figure}[h]
		\centering
		\includegraphics[width=.49\textwidth]{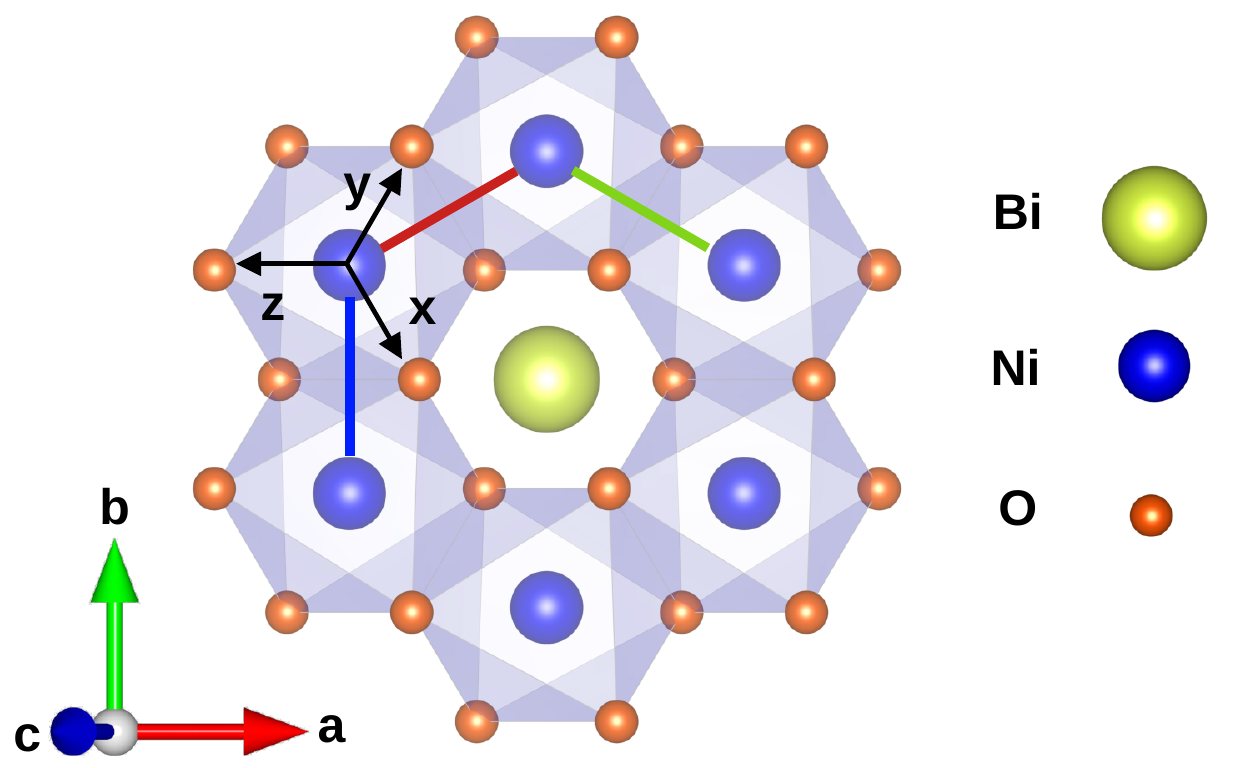}
		\caption{Projection on the $ab$ plane of the crystal structure  of Na$_3$Ni$_2$BiO$_6$. Each honeycomb layer consists of Ni atoms surrounded by oxygen octahedra. The center of each honeycomb is occupied by a single Bi atom. The axes of the local Wannierization coordinate system \textbf{x}, \textbf{y} and \textbf{z} are oriented along the bonds between Ni and O atoms. The \textbf{X} (red), \textbf{Y} (green) and \textbf{Z} (blue) Ni-Ni bonds are depicted inside the octahedral structure.}
		\label{fig:Structure}
	\end{figure}

	\section{RESULTS}

	\subsection{Magnetic Exchange Couplings}
	\vspace{0.2cm}

	Magnetic exchange couplings are obtained through the {\it ab-initio}-based projED method~\cite{riedl2019ab,kaib2022electronic,konieczna2025understanding} discussed in detail in the \textit{Methods} section. We calculated all magnetic couplings, up to the 3\textsuperscript{rd} NN. We include all symmetry-allowed bilinear exchanges, the isotropic biquadratic exchanges, as well as the SIA. The resulting magnetic Hamiltonian is
	\begin{align}
		\mathcal{H} = \mathcal{H}_{\rm SIA} + \mathcal{H}_1 + \mathcal{H}_2 + \mathcal{H}_3,
	\end{align}
	where $\mathcal{H}_{\rm SIA}$ is given by 
	\begin{align}\label{eq:HamSIA}
		\mathcal{H}_{\rm SIA} = \sum_i \sum_a S_i^a \Lambda^a S_i^a,
	\end{align}
	and $\mathcal{H}_{n}$ is the $n$\textsuperscript{th} NN interaction term, given by
	\begin{align}\label{eq:biquadraticHamiltonian}
		\mathcal{H}_n = \sum_{\braket{i,j}_n} \left(\sum_{a,b} S_i^aJ_{ij}^{ab}S_j^b\right) + B_{ij}(S_iS_j)^2.
	\end{align}
    Here, $J^{ab}_{ij}$ is the bilinear exchange matrix on bond $\braket{i,j}_n$, coupling $a$, $b$ spin components on site $i$ and $j$ respectively, and $B_{ij}$ is the isotropic biquadratic exchange. In the cubic \textbf{xyz} basis, whose axes point approximately towards the neighboring ligands (see~\cref{fig:Structure}), the bilinear exchange matrix on the $Z$-bonds takes the symmetry-reduced form
	\begin{align}
		J_{n=1,3}^Z=\begin{pmatrix}
			J_n & \Gamma_n & \Gamma'_n \\
			\Gamma_n & J_n & \Gamma'_n \\
			\Gamma'_n & \Gamma'_n & J_n+K_n
		\end{pmatrix}
	\end{align}
	on the 1\textsuperscript{st} and 3\textsuperscript{rd} NN, and the form
	\begin{align}
		J_{2}^Z=\begin{pmatrix}
			J_2 & \Gamma_2+D^z & \Gamma'_2-D^y \\
			\Gamma_2-D^z & J_2 & \Gamma'_2+D^x \\
			\Gamma'_2+D^y & \Gamma'_2-D^x & J_2+K_2
		\end{pmatrix}
	\end{align}
	on the 2\textsuperscript{nd} NN. 

    In the fully $C_3$-symmetric model, the exchange matrices for the $X$-, $Y$-, and $Z-$bonds are related by symmetry. In the NNBO case, however, $C_3$ symmetry is broken. Although the $X$- and $Y$-bond couplings remain related by a mirror symmetry, the two sets of parameters, one for $X$- and $Y$-bonds, and one for $Z$-bonds, must be distinguished.
    
    The extended Hubbard Hamiltonian used in the numerical derivation of the exchange parameters  features an additional effective nearest-neighbor superexchange term of 4\textsuperscript{th} order in perturbation theory (see {\it Methods} section). It captures intersite Coulomb interactions between metal and ligand atoms that are not adequately described by standard DFT. This contribution is relevant in systems with strong metal-ligand hybridization, as is the case in NNBO. This effective exchange is controlled by a screening parameter $\gamma$, which, in the context of this model, serves as a fitting parameter (see \textit{Methods} section for details). 

    \begin{table}
		\begin{tabular}{c|cc|cc|cc}
							& \multicolumn{2}{c|}{$n=1$} & \multicolumn{2}{c|}{$n=2$} & \multicolumn{2}{c}{$n=3$} \\
							& $\eta=X$	& $\eta=Z$	& $\eta=X$	& $\eta=Z$	& $\eta=X$	& $\eta=Z$	\\
			\hline
			$J_n^\eta$		& -3.643	& -3.775	&  0.057	&  0.048	&  1.29		&  1.517	\\
			$K_n^\eta$		& -0.018	& -0.016	&  0.003	&  0.002	& -0.018	& -0.017	\\
			$\Gamma_n^\eta$	& -0.004	& -0.004	&  0.002	&  0.002	& -0.001	& -0.001 	\\
			$\Gamma_n^\eta$	&  0.003	&  0.003	& -0.001	& -0.001	&  0		&  0  		\\
			$D^x$			&  -		& -			& -0.010	&  0.004	& -			& -			\\
			$D^y$			&  -		& -			&  0.007	&  0.004	& -			& -			\\
			$D^z$			&  -		& -			&  0.007	&  -0.008	& -			& -			\\
			$B_n$			&  0.016	&  0.012	&  0		&  0		& -0.002	& -0.003	\\
		\end{tabular}
		\caption{Resulting couplings, in units of meV, from projED, using $\gamma=0.28$. Listed are the bilinear magnetic couplings, in the cubic \textbf{xyz} basis, including Heisenberg $J_n$, Kitaev $K_n$, symmetric off-diagonal $\Gamma_n$ and $\Gamma'_n$, as well as the biquadratic exchange $B$, for the first three NN bonds. Results are partitioned on the two symmetry inequivalent sets of $X$($Y$)- and $Z$-bonds, with only small differences observed due to the near $C_3$-symmetric structure.}
		\label{tab:MagCouplings}
	\end{table}
    
    ~\cref{tab:MagCouplings} shows the results for the magnetic exchange couplings obtained with $\gamma=0.28$. In order to set this screening parameter, we compare the result to a perturbative estimate based on the derivations in~\cite{stavropoulos2021emergent}. We note that, in the perturbative treatment, analytical tractability was achieved by controlling the number of expansion terms through the use of an artificially higher-symmetry nearest-neighbor hopping model, in which most small hoppings were set to zero. In addition, the expansion was restricted to processes that only involve the metal $e_g$ orbitals. As a result, specific material details and additional nearest-neighbor terms -- captured in the  projED method -- are not included. Nevertheless, the perturbative approach for nearest neighbor interactions still provides reliable order-of-magnitude estimates, against which we can benchmark  $\gamma$. The perturbative estimate $J_1^{\text{pert}}\simeq-3.57\text{ meV}$ is in good agreement with the projED ($\gamma=0.28$) results for $J_1 = (J_1^X + J_1^Y + J_1^Z)/3 = -3.687\text{ meV}$. If we completely neglect the effective NN superexchange, effectively setting $\gamma=0$, we would get $J_1 = 0.464\text{ meV}$ from projED, in complete disagreement to the perturbative estimation. 
    
	We observe that the couplings (\cref{tab:MagCouplings}) are dominated by the 1\textsuperscript{st} and 3\textsuperscript{rd} NN Heisenberg exchanges $J_1$ and $J_3$. The corresponding Kitaev terms $K_{1,3}$ as well as 2\textsuperscript{nd} NN $J_2$ have only small contributions. All additional off-diagonal terms including symmetric off-diagonal $\Gamma$, $\Gamma'$ and Dzyaloshinskii-Moriya terms $D$ are of the order of $10^{-3}$--$10^{-2}$meV. The biquadratic terms $B$ are found to be of a non-negligible but small magnitude. Placing the terms in a hierarchy of energies, the next term after $J_1$ and $J_3$ is actually the SIA which we now analyze in detail.

	\subsection{Single-ion anisotropy and magnetic moment}
	\vspace{0.2cm}
	
	The SIA coupling matrix in~\cref{eq:HamSIA} is found to be
	\begin{align}
		\Lambda = \begin{pmatrix}
			0.206 & -0.285 & -0.248 \\
			-0.285 & 0.206 & -0.248 \\
			-0.248 & -0.248 & 0.260
		\end{pmatrix}\ \text{meV}.
	\end{align}
    Specifically for the SIA term it is helpful to transform from the cubic \textbf{xyz} basis, to the crystallographic $\mathbf{abc^*}$ coordinates, with $\mathbf{c^*}$ perpendicular to the $ab$-plane (see the \textit{Methods} section for details). The transformed matrix reads
	\begin{align}\label{eq:SIA}
		\tilde{\Lambda} = \begin{pmatrix}
			0.477 & 0 & -0.043 \\
			0 & 0.491 & 0 \\
			-0.043 & 0 & -0.297
		\end{pmatrix}\ \text{meV}.
	\end{align}
    In this basis, the preference of the SIA for the $\mathbf{c^*}$ axis is visible. Neglecting the small off-diagonal term, and noting that the diagonal elements of the SIA encode the spin-length constraint -- allowing a constant to be subtracted from the diagonal -- we conclude that the SIA contributes approximately a value $ -0.781 (S^{c^*}_i)^2$ on each site. This makes it the next most important energy scale after $J_1$ and $J_3$. For a more accurate analysis of the preferred direction, in ~\cref{tab:SIA_Axes} we display the energies and eigenvectors corresponding to $\tilde{\Lambda}$. We find that the easy axis selected by the SIA lies approximately along the vector $\left( 0,\ 0,\ 1 \right)$ in the $\mathbf{abc^*}$ coordinates.
	\begin{table}
		\begin{tabular}{c|c}
			Energies (meV) & Axes direction \\
			\hline
			-0.299 & $\left( 0.055,\ 0,\ 1 \right)$ \\
			 0.480 & $\left( 1,\ 0,\ -0.055 \right)$ \\
			 0.491 &  $\left( 0,\ 1,\ 0 \right)$
		\end{tabular}
		\caption{Energy eigenvalues, in units of meV, and corresponding normalized eigendirections from the SIA coupling matrix in~\cref{eq:SIA} in $\mathbf{abc^*}$ coordinates. The easy axis of Ni$^{2+}$ atoms is oriented (mostly) along the $\mathbf{c^*}$ direction.}
        \label{tab:SIA_Axes}
	\end{table}

    \subsection{Ground state characterization}
	\vspace{0.2cm}

	\begin{figure*}[t]
 		\centering
       \begin{overpic}[width=1.0\textwidth,percent,grid=false,tics=2]{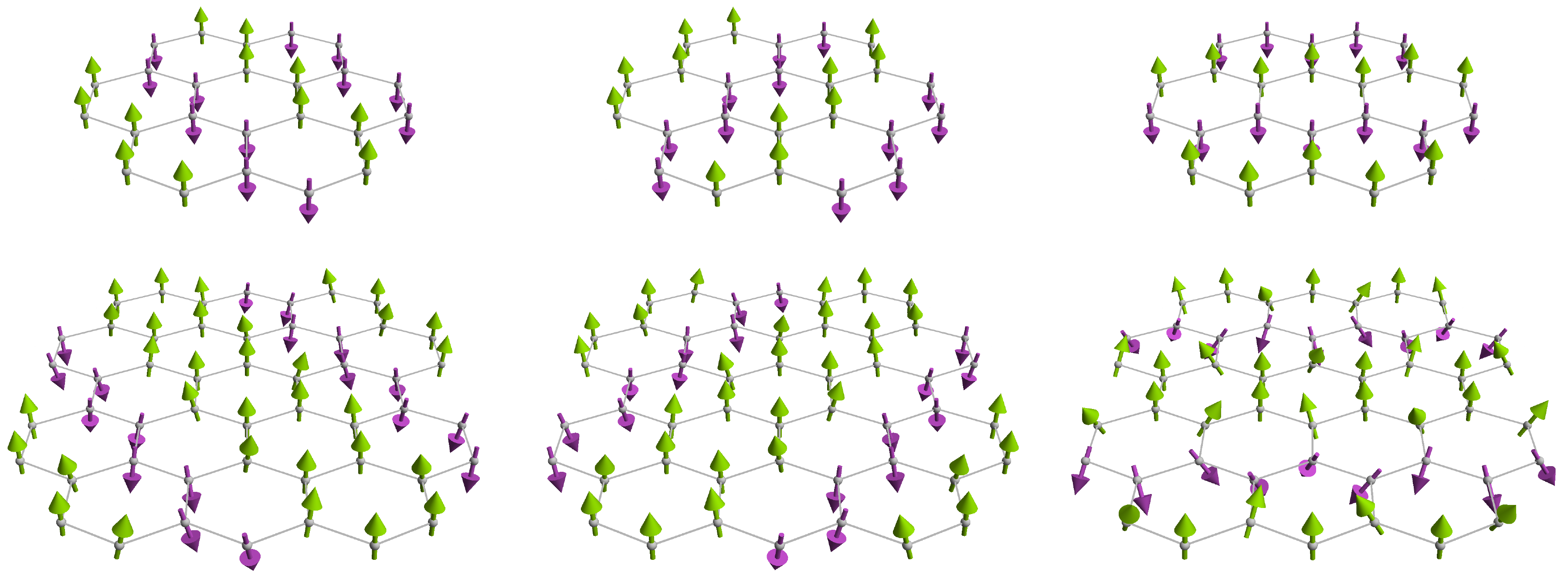}
        \put(0,35){\textbf{a}}
        \put(33.3333,35){\textbf{b}}
        \put(66.6666,35){\textbf{c}}
        \put(0,19){\textbf{d}}
        \put(33.3333,19){\textbf{e}}
        \put(66.6666,19){\textbf{f}}
       \end{overpic}  
		\caption{Ground state spin configuration in zero-field (\textbf{a}-\textbf{c}) and $\mu_0H=5.5$T, $H\vert\vert c^*$ (\textbf{d}-\textbf{f}). In the zero-field environment, three distinct zigzag phases in close energetic proximity (\textbf{c} higher in energy by $\sim0.1$meV) to one another coexist in different honeycomb layers and domains. Similarly, for $\mu_0H=5.5$T three distinct 1/3-magnetization phases in close energetic proximity coexist (\textbf{f} higher in energy by $\sim0.03$meV)}
		\label{fig:SpinConfiguration}
	\end{figure*}
	\begin{figure}[h]
		\centering
        \includegraphics[width=.48\textwidth]{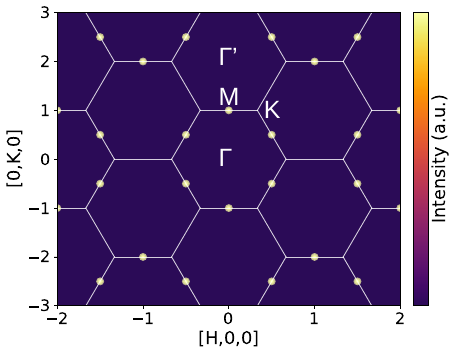}
		\caption{Static structure factor $S(\textbf{q})$ at zero magnetic field corresponding to the three possible ground state spin configurations (see~\cref{fig:SpinConfiguration}~\textbf{a}-\textbf{c}). Data is presented for the $(H,0,0)$, $(0,K,0)$ momentum space plane with $H$ and $K$ in terms of fractional \textbf{k}-space lattice units. The displayed spectrum consists of the averaged sum of three spectra that each result from one of the three different ground state spin configurations. The data for the individual configurations is visible in Supplementary Note 3,~\cref{Supp:fig:ENS_NoField}.}
		\label{fig:ENS_NoField}
	\end{figure}
	\begin{figure}[h]
		\centering
		\includegraphics[width=.48\textwidth]{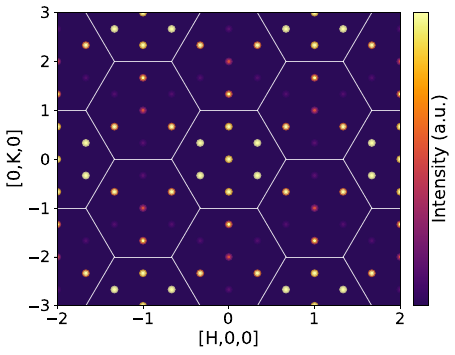}
		\caption{Static structure factor $S(\textbf{q})$  at a magnetic field $\mu_0H=5.5$T, $H\vert\vert c^*$ for the three possible ground state spin configurations (see~\cref{fig:SpinConfiguration}~\textbf{d}-\textbf{f}). Data is presented for the $(H,0,0)$, $(0,K,0)$ momentum space plane with $H$ and $K$ in terms of fractional \textbf{k}-space lattice units. The displayed spectrum consists of the averaged sum of three spectra that each result from one of the three different ground state spin configurations. The data for the individual configurations is visible in Supplementary Note 3,~\cref{Supp:fig:ENS_Field}.}
		\label{fig:ENS_Field}
	\end{figure}
    Spin configurations representing the ground state at zero and finite temperature are obtained through use of the Su(n)ny \cite{dahlbom2025sunny} software. We employ the Monte Carlo procedure to determine the ground state spin configuration, with and without an external magnetic field of $\mu_0H=5.5$T, applied perpendicular to the honeycomb plane. Results were obtained on a 18x18x1 primitive lattice for a target temperature of $0.001$K. 
    
    In the zero-field case, the model stabilizes a zigzag ground state, characterized by a propagation vector along the $X$ or $Y$-bonds, as shown in~\cref{fig:SpinConfiguration}~\textbf{a} and \textbf{b}. Due to the slightly broken $C_3$ symmetry, a zigzag configuration with $Z$-bond propagation, shown in~\cref{fig:SpinConfiguration}~\textbf{c}, lies $\Delta E_{gs}=0.1$meV higher in energy. This is consistent with the model parameter hierarchy $|J_1^Z|>|J_1^{X(Y)}|$, since a ferromagnetic coupling on the $Z$-bond incurs a higher energy cost for the anti-aligned moments in a $Z$-bond--propagating zigzag state than in those propagating along the $X$ or $Y$-bonds. 
    
    Overall, the spin moments are pinned to the $\mathbf{c}^*$ axis, due to the SIA easy-axis selection. The slight energy difference in the zigzag propagations corresponds to a temperature scale of $\sim1$K, allowing all three configurations to coexist in a real material at finite temperature in separate domains or layers. Next, we compute the static structure factor (SSF) $\mathcal{S}(\mathbf{q})$ from the spin configurations discussed above. It is evaluated for all allowed momentum points on a two-dimensional slice along the $(H,0,0)$ and $(0,K,0)$ directions, corresponding to reciprocal-lattice directions of the conventional unit cell. $H\in [-2,2]$ and $K\in [-3,3]$ in fractional units, in order to compare to the ENS data in~\cite{shangguan2023onethirdmag}. In a first step, the SSF is obtained separately for all three configurations~\cref{fig:SpinConfiguration}~\textbf{a}-\textbf{c}. Assuming any real measurement would sample all three realizations, the resulting data is averaged and the final result displayed in~\cref{fig:ENS_NoField}. We find that the $\mathbf{M}$ point correlators are peaked, consistent with the appearance of the zigzag order. 
    
    Having understood the zero field state, we move on to analyze the $\mu_0H=5.5$T field state. We again find a similar behavior, with two ground states propagating along $X$- or $Y$-bond directions and related by mirror symmetry, as shown in~\cref{fig:SpinConfiguration}~\textbf{a} and \textbf{b}. Additionally, again a third $Z$-bond propagating state with a slightly higher energy by $0.03$meV is found and visible in~\cref{fig:SpinConfiguration}~\textbf{c}. 
    
    All three propagations are characterized by a one-third magnetization, realized by a $\uparrow\uparrow\downarrow$ spin alignment. Here, islands of two aligned chains interchange with one anti-aligned chain. In the following, we will refer to this configuration as the 'double zigzag' state. As before, due to the small energy difference, we average the static structure factor of the three double zigzags and present this result in~\cref{fig:ENS_Field}. The peak positions align with the ENS data in~\cite{shangguan2023onethirdmag}. We observe dominant peaks at the $\frac{2}{3}$\textbf{M} points, consistent with the double zigzag $\uparrow\uparrow\downarrow$-spin-chain pattern. There is also a significant $\mathbf{\Gamma}$ point contribution, since the double zigzag state has finite magnetization. 
    
    We note that there are also very fine satellite features nearby the dominant peaks, which are too small to be observed in the color scale. Also, system scaling trials across a range of symmetric $L\times L\times 1$ clusters showed a slight gradual lowering of the ground state energy with increasing system size $L$, on average of the order of $\sim10^{-4}$ meV for $L\rightarrow L+1$. This is indicative that the true ground state in the thermodynamic limit has some potentially incommensurate fine features, which we cannot capture in the finite size calculations. 
    
    Finally, note that the double zigzag $\uparrow\uparrow\downarrow$-spin-chain pattern described above naturally gives rise to a magnetization equal to one third of the fully saturated polarized state.
    To further elucidate the origin of the $1/3$-magnetization phase, we next analyze the systems magnetization behavior, focusing on the extent and stability of this phase.

	\subsection{Magnetization}
	\vspace{0.2cm}

	\begin{figure}[h]
		\centering
		\includegraphics[scale=.38]{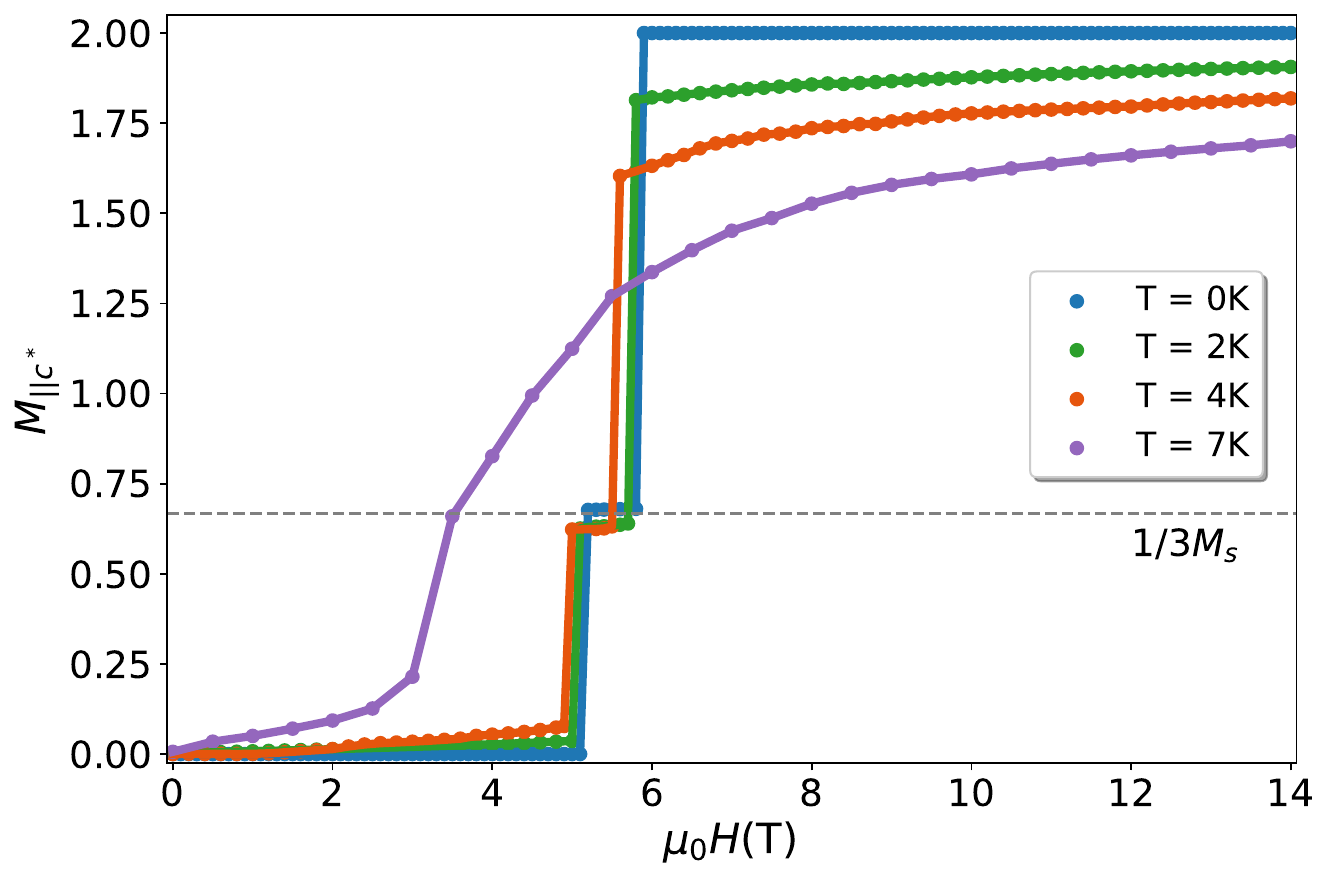}
		\caption{Magnetization curve of NNBO, representing the magnetization parallel to $c*$ per Ni$^{2+}$, for different values of temperature T, for fixed $\gamma=0.28$. The 1/3-plateau phase is observed for $0$, $2$ and $4$K, while at $7$K the intermediate phase disappears.}
		\label{fig:MagnetizationCurve_Temperature}
	\end{figure}
    \begin{figure}[t]
 		\centering
       \begin{overpic}[width=0.48\textwidth,percent,grid=false,tics=2]{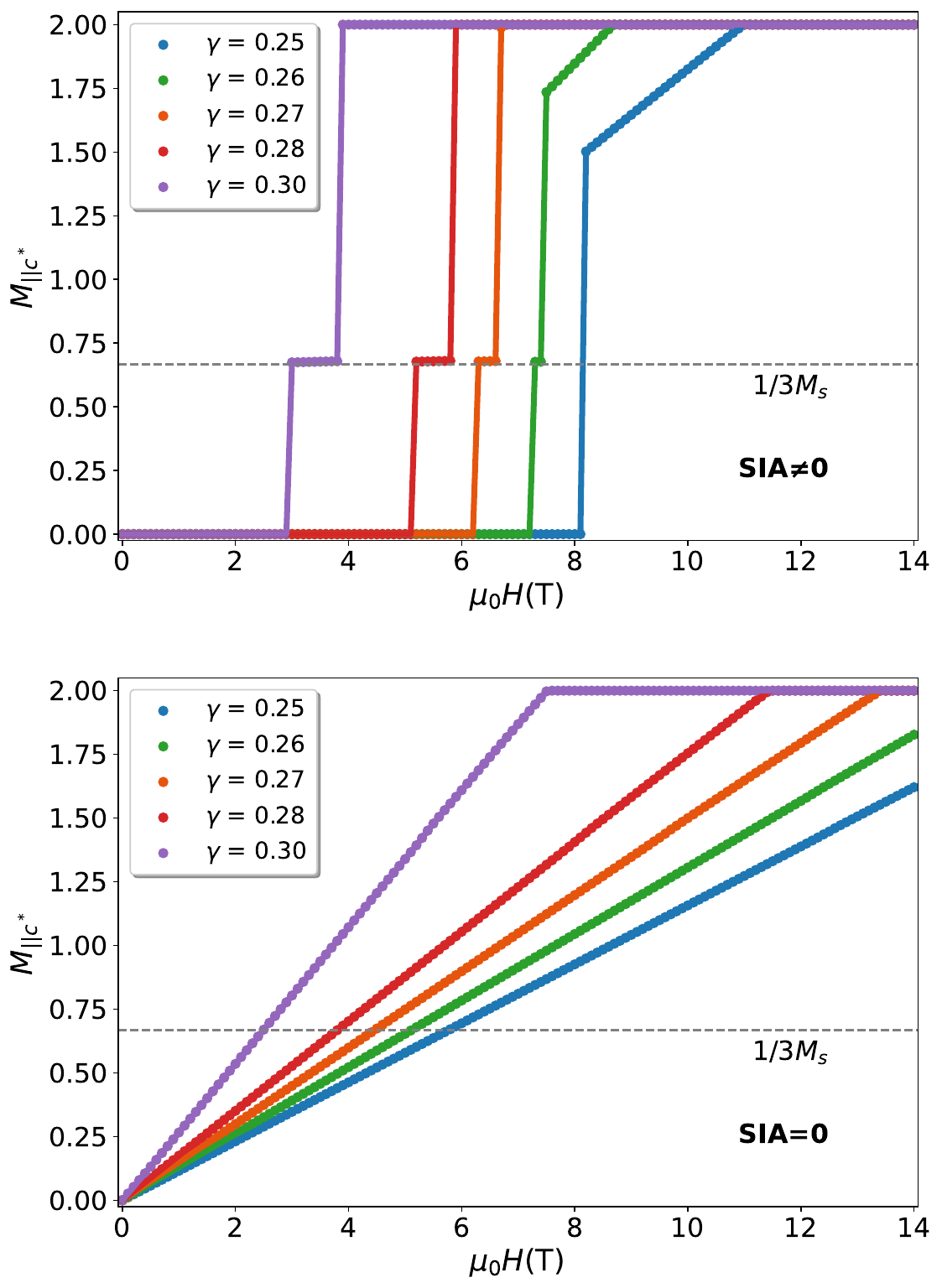}
        \put(-1,99){\textbf{a}}
        \put(-1,49){\textbf{b}}
        \end{overpic}  
	    \caption{\textbf{a} $T=0$K magnetization curves of NNBO, representing the magnetization parallel to $c*$ per Ni$^{2+}$, for different values of the screening parameter $\gamma$, which change the exchange couplings. The 1/3 plateau phase has a finite region of appearance for $\gamma$ starting at $\geq0.26$ until at least $0.30$ \textbf{b} The same plot where SIA term has been removed. The 1/3-plateau phase completely vanishes.}
		\label{fig:MagnetizationCurve_Gamma}
	\end{figure}
    We start by investigating the $1/3$-magnetization plateau as a function of temperature $T$ and its sensitivity to the screening parameter $\gamma$. To this end, we apply Landau-Lifshitz spin dynamics in Su(n)ny by thermalizing the system for each value of $\mu_0H$ based on the ground state configurations corresponding to the data in~\cref{fig:MagnetizationCurve_Gamma} for $\gamma=0.28$. 
    
    In~\cref{fig:MagnetizationCurve_Temperature} we show the magnetization curve as a function of magnetic field $\mu_0H$ applied along $\mathbf{c^*}$, for exchanges calculated from a fixed screening parameter $\gamma=0.28$ at various temperatures. At high temperature ($T=7$~K), no plateau phase survives, and the system becomes fully polarized at high fields. As the temperature decreases, an intermediate $1/3$-magnetization plateau emerges, slightly reduced below $1/3$ by thermal fluctuations, and persists down to $T=0$, where the plateau reaches its maximum extent with fully saturated $1/3$ magnetization. 
    
    The 1/3-plateau phase at $T=0$ extends between $\mu_0H=5.0$T and $5.9$T. This region is slightly below what is observed experimentally \cite{shangguan2023onethirdmag}. However, some degree of tuning is possible through the screening parameter $\gamma$. Varying $\gamma$ primarily  affects the first-neighbor Heisenberg coupling $J_1$, while $K_1$, $\Gamma_1$, and $\Gamma_1'$ change only slightly within the same order of magnitude. Thus, adjusting $\gamma$ effectively tunes the ferromagnetic $J_1$, which is minimal at $\gamma=0$ ($J_1^Z=-0.469$) and increases with $\gamma$. The resulting $1/3$-plateau phases for different $\gamma$ values are shown in~\cref{fig:MagnetizationCurve_Gamma}~\textbf{a}.
    If the screening is chosen too small, $\gamma\leq0.25$, the 1/3-plateau phase vanishes. For $\gamma=0.25$ up to a field of $8.1$T the zero-field zigzag phase is sustained. At larger fields the system switches to a ferromagnetic state with a canting angle with respect to $c^*$, which subsequently gets smaller while further increasing $\mu_0H$. Increasing $\gamma$ at first introduces a tiny plateau region, that increases subsequently as $\gamma$ continues getting larger. Simultaneously, the plateau region overall moves to lower magnetic fields. We find that $0.27\leq\gamma\leq0.28$ is the best choice, producing a large enough plateau at fields close to the experimentally observed range. 

    We also highlight how crucial SIA is for the stabilization of the phase. If we take the same exchange parameters used in~\cref{fig:MagnetizationCurve_Gamma}~\textbf{a} and remove the SIA term, the plateau phase disappears entirely, as seen in panel \textbf{b} of the figure.

	\section{DISCUSSION}
	
	\vspace{0.2cm}
	
	The magnetic exchange couplings presented in~\cref{tab:MagCouplings} are obtained by combining DFT and the projED method including the single fitting parameter $\gamma$. These results suggest that pure Heisenberg exchange between 1\textsuperscript{st} and 3\textsuperscript{rd} NN bonds, as well as a large SIA contribution, are the driving mechanisms for the magnetic properties in NNBO. Furthermore, the evaluation of the SSF and field-dependent magnetization based on these exchanges closely match the experimental results in~\cite{shangguan2023onethirdmag}.     
    Including the screening parameter $\gamma$, which is an effective nearest-neighbor superexchange term of 4\textsuperscript{th} order in perturbation theory, as was done in~\cite{konieczna2025understanding}, is crucial for arriving at these exchanges, and subsequently getting the correct description of the magnetic properties of NNBO.
    
     The presence of SIA is crucial for the occurrence of the 1/3 plateau, as we show that without it the plateau phase disappears at all field values. However, anisotropic couplings like Kitaev terms play only a minor role. This is because mechanisms that typically generate substantial Kitaev couplings, such as spin-orbit interactions -- crucial in e.g. $\alpha$-RuCl$_3$, are weak in NNBO. Large distortions of the octahedral oxygen structure, which can lead to Ising type anisotropies on ligand-mediated exchanges, as present for example in CoNb$_2$O$_6$~\cite{Coldea_2023,konieczna2025understanding,churchill2024}, are also absent. 
    
    The model realizes the ground state spin configurations, in zero-field and at intermediate magnetic fields, visible in~\cref{fig:SpinConfiguration}. We were able to show that the phase at intermediate fields consists of two degenerate ground states and one energetically close state at $\Delta E_{gs}=0.03\ \text{meV}$ which exhibit a $\uparrow\uparrow\downarrow$ double zigzag configuration along three different propagation directions. This result is a consequence of the slightly broken rotational $C_3$ symmetry with $|J_1^Z|>|J_1^{X,Y}|$, leading to preferred propagations along the $X$- and $Y$-bonds, while the $Z$-bond propagation corresponds to an energetically higher state.

    We conclude that the $1/3$-magnetization plateau arises from a compromise between strong ferromagnetic $J_1$ and antiferromagnetic $J_3$ couplings, stabilized by a large SIA that keeps spins orthogonal to the $ab$-plane. Under this constraint, the 1/3-plateau state minimizes the energy, with two aligned chains forming a small ferromagnetic region alternating with an anti-aligned chain, producing a ground state that is partially ferro- and antiferromagnetic, satisfying the competing $J_1$ and $J_3$ interactions.

    \vspace{0.2cm}
	\section{METHODS}
	\vspace{0.2cm}
	\subsection{Structure and Symmetries}
	
	\vspace{0.2cm}
	
	All calculations performed in this work are based on the experimental structure data presented in Ref.~\cite{seibel2013Structure}. Na$_3$Ni$_2$BiO$_6$ crystallizes in the space group $C2/m$. The system exhibits a quasi-two-dimensional layered Ni-based honeycomb structure in the crystallographic $ab$-plane. The $c$-axis has an angle of $108.56^\circ$ to $a$ and $90^\circ$ to $b$. For further considerations, we define $c^*$ as the direction perpendicular to the $ab$-plane, tilted by $18.56^\circ$ with respect to $c$. As visible in~\cref{fig:Structure} the spin-1 Ni$^{2+}$ ions are surrounded by edge-sharing oxygen octahedra. Single Bi atoms are located at the center of each honeycomb unit while Na atoms can be found in between the layers.
    
    Na$_3$Ni$_2$BiO$_6$ displays magnetic ordering below $T_n\sim11$K with antiferromagnetic zigzag chains along the crystallographic $a$-direction~\cite{seibel2013Structure, shangguan2023onethirdmag}. Within the honeycomb structure, the $C_3$ rotational symmetry between Ni bonds is slightly broken, leading to a distinct $Z$-bond, while the $X$ and $Y$-bonds are related by a mirror plane perpendicular to $a$. Therefore, in this work, magnetic coupling data will be distinguished between the $X$ and $Z$-bonds.

    \subsection{Density Functional Theory}

	\vspace{0.2cm}

    The aim of obtaining the magnetic couplings of NNBO, which are presented in this work, requires, in a very first step, the computation of magnetically relevant hoppings and crystal field data of the compound. This data can be generated by means of DFT calculations based on the material's structure. The computation was performed using the Full Potential Local orbital (FPLO) basis ~\cite{koepernik1999full}, within the Generalized Gradient approximation (GGA)~\cite{perdew1996generalized,eschrig2003density} in a full-relativistic, non-magnetic setup on a 12x12x12 grid. The relativistic calculation serves to additionally compute information on the spin-orbit coupling. 
    
    The computation of hopping parameters in FPLO requires the choice of a  local Wannierization basis for the construction of appropriate Wannier functions~\cite{koepernik2023symmetry}. In NNBO, the functions are constructed for the Ni 3$d$ orbitals and the cubic basis \textbf{x}, \textbf{y} and \textbf{z} is chosen with axes aligned approximately along the metal-ligand Ni-O bonds within the octahedral structure. The Wannier basis in terms of the FPLO coordinate system $a^*$, $b$ and $c$ ($a^*$ in the $ac$-plane, tilted by $18.56^\circ$ from $a$) is
	\begin{align}\label{eq:CubicAxes1}
		\textbf{x}=\begin{pmatrix}
			0.5708 \\
			-1/\sqrt{2} \\
			0.4174
		\end{pmatrix},\
		\textbf{y}=\begin{pmatrix}
			0.5708 \\
			1/\sqrt{2} \\
			0.4174
		\end{pmatrix},\
		\textbf{z}=\begin{pmatrix}
			-0.5902 \\
			0 \\
			0.8072
		\end{pmatrix},
	\end{align}
	while in the global $abc^*$-basis, where $ab$ is the honeycomb plane and $c^*$ the direction perpendicular to it, the Wannier basis is written as
	\begin{align}\label{eq:CubicAxes2}
		\textbf{x}=\begin{pmatrix}
			0.4082 \\
			-1/\sqrt{2} \\
			0.5774
		\end{pmatrix},\
		\textbf{y}=\begin{pmatrix}
			0.4082 \\
			1/\sqrt{2} \\
			0.5774
		\end{pmatrix},\
		\textbf{z}=\begin{pmatrix}
			-0.8165 \\
			0 \\
			0.5774
		\end{pmatrix}.
	\end{align}

	\vspace{0.2cm}
	\subsection{Multi-Orbital Hubbard Model}
	
	\vspace{0.2cm}

    The calculation of hopping parameters from DFT, as described above, is the first step of the \textit{ab-initio} based method projED \cite{riedl2019ab}, which ultimately leads to the computation of the magnetic couplings. 
    
    In a following step, these parameters serve as input for a multi-orbital Hubbard model for the Ni 3$d$ orbitals. Exact diagonalization of the resulting Hamiltonian on finite-size clusters yields the states relevant for the ground state configuration, which is then projected from the orbital basis into an effective pseudo-spin basis, giving us the magnetic coupling Hamiltonian between two sites. For the system in question, the Ni $3d$ orbitals are occupied by 8 electrons with ground state configuration $\left(t_{2g}\right)^6\left(e_g\right)^2$. Since the $t_{2g}$ levels are fully occupied it suffices to project onto the $e_g$ orbitals where the remaining electrons form an effective $S=1$ state. Details on the projection can be found below. 
    
	The aforementioned Hubbard-Hamiltonian in the basis of Ni $3d$ orbitals consists of the following terms
	\begin{align}\label{eq:HubbardHamiltonian}
		\mathcal{H} = \mathcal{H}_{\rm hop} + \mathcal{H}_{\rm nn-U} + \sum_i\mathcal{H}_i.
	\end{align}
	In~\cref{eq:HubbardHamiltonian} $\mathcal{H}_{\rm hop}$ describes the hopping processes between different $3d$ orbitals, where the hopping integrals were obtained from an FPLO calculation. It can be written in the form
	\begin{align}
		\begin{aligned}
			\mathcal{H}_{\rm hop} = \sum_{\braket{ij}}\sum_{ab}\sum_{\sigma\sigma'}t_{ij,\sigma\sigma'}^{ab}d_{ia\sigma}^\dagger d_{jb\sigma'}.
		\end{aligned}
	\end{align}
	$\mathcal{H}_{\rm nn-U}$ is an effective Coulomb interaction term between Ni NN atoms which originates from fourth order superexchange processes that are not explicitly accounted for within DFT. The corresponding term can be written as
	\begin{align}
		\begin{aligned}
			\mathcal{H}_{\rm nn-U} = \gamma\sum_{\braket{ij}}\sum_{ab} \tilde{J}\Big[ & \sum_{\sigma} \frac{n_{ia\sigma}n_{jb\sigma}}{4} - \frac{n_{ia\uparrow}n_{jb\downarrow}}{2} \\
			& + d_{ia\uparrow}^\dagger d_{jb\downarrow}^\dagger d_{ia\downarrow}d_{jb\uparrow} \Big].
		\end{aligned}
	\end{align}
	The strength of this interaction is computed from FPLO Wannier function overlaps between Ni $3d$ and oxygen $2p$ orbitals. It is incorporated in the coupling matrix $\tilde{J}$, whose full expression is given in Supplementary Note 1. Since in the ground state configuration of NNBO the $t_{2g}$ levels are fully occupied, the interaction of the $e_g$ orbitals with the ligands plays a dominant role in the resulting couplings. This is reflected in the size of the overlap the $x^2-y^2$ orbitals have with the ligands (see Supplementary Note 1). A detailed discussion on the relevance of $e_g$ orbital hybridization with ligands in edge-sharing octahedral structures can be found in~\cite{autieri2022limited}. The presence of the prefactor $\gamma$ in the Hamiltonian term $\mathcal{H}_{\rm nn-U}$ originates in the necessary screening of the process, due to its overestimation within FPLO. A detailed discussion of the implementation of $\mathcal{H}_{\rm nn-U}$ can be found in~\cite{konieczna2025understanding}. 
    
	The last term of the onsite Hubbard Hamiltonian $\mathcal{H}_i$ consists of
	\begin{align}
		\mathcal{H}_i = \mathcal{H}_{\rm CF} + \mathcal{H}_{\rm SOC} + \mathcal{H}_{\rm U},
	\end{align}
	which includes crystal field splitting effects, spin-orbit coupling and Coulomb interaction. All spin-orbit coupling effects as well as crystal field splitting is accounted for from the FPLO calculation. The Coulomb term is included in its most general form
	\begin{align}
		\mathcal{H}_{\rm U} = \frac{1}{2} \sum_{\sigma\sigma'}\sum_{abcd} U_{abcd}d_{ia\sigma}^\dagger d_{ib\sigma}^\dagger d_{ic\sigma}d_{id\sigma}
	\end{align}
	with $3d$ orbitals $a$, $b$, $c$ and $d$ and spin orientations $\sigma$/$\sigma'$. The matrix elements $U_{abcd}$ are fully determined from the parameters $U_{avg}=5.5$eV and $J_{avg}=0.15U_{avg}$, as well as Gaunt coefficients (see~\cite{pavarini2011lda}), which result from $d$ orbital spherical harmonics. The choice of  $U_{avg}$ and $J_{avg}$ is based on the discussion in~\cite{zvereva2015zigzag} of appropriate Coulomb interaction strengths in Ni-based honeycomb monolayers as well as the values used in~\cite{shangguan2023onethirdmag}.

	\vspace{0.2cm}
	\subsection{The $S_{\rm eff}=1$ ground state}
	
	\vspace{0.2cm}
	
	After the Hubbard Hamiltonian (see~\cref{eq:HubbardHamiltonian}), constructed following the procedure above, is exactly diagonalized one can extract the resulting ground state energies and eigenstates in orbital basis. In the following step a projection onto the spin-$1$ basis has to be performed. Since the $t_{2g}$ levels are fully occupied only the half-filled $e_g$ orbitals need to be taken into account. Since on each of the two $e_g$ levels an electron can be either in a $\ket{\uparrow}$ or $\ket{\downarrow}$ state, a total of three states with a spin of $1$ can be assigned to each Ni$^{2+}$ ion:
	\begin{subequations}\label{eq:Spin1States}
		\begin{align}
			\ket{1,1}=&\ket{\uparrow\uparrow} \\
			\ket{1,0}=&\frac{1}{\sqrt{2}} \left(\ket{\uparrow\downarrow}+\ket{\downarrow\uparrow}\right) \\
			\ket{1,-1}=&\ket{\downarrow\downarrow}
		\end{align}
	\end{subequations}
	In order to cast the ground state eigenstates in orbital basis to the spin-$1$ basis the vectors are projected onto the two-site product space of the single-site states in~\cref{eq:Spin1States}. The resulting two-site $9x9$ Hamiltonian for the spin-$1$ states is then, in a subsequent step, projected onto the final pseudo-spin Hamiltonian as presented in~\cref{eq:biquadraticHamiltonian} with magnetic couplings $J_{ij}^{ab}$ between two sites $i$ and $j$ for $a,b=x,y,z$.

	\vspace{0.2cm}
	\section{ACKNOWLEDGEMENTS}
	
	\vspace{0.2cm}

	We acknowledge support by the Deutsche Forschungsgemeinschaft (DFG, German Research Foundation) for funding through project TRR 288 — 422213477 (project A05) and through QUAST-FOR5249 - 449872909 (project TP4).

	\vspace{0.2cm}
	\section{COMPETING INTERESTS}
	
	\vspace{0.2cm}
	
	The authors declare no competing interests.
	
	\clearpage

	\appendix

	\section{Supplementary Note 1: Nearest Neighbor Coulomb}
	
	\vspace{0.2cm}
	
	The effective Coulomb interaction strength between two neighboring Ni atoms are incorporated in the matrix
	\begin{align}\label{Supp:eq:nnCoulombMatrix}
		\tilde J = J_HA_{ab} + (U_0-J_H) B_{ab} + U_0C_{ab},
	\end{align}
	with
	\begin{subequations}\label{Supp:eq:WannierOverlap}
		\begin{align}
			A_{ab} \equiv & \gamma\sum_{n,\alpha\neq\beta} |\phi_{ia}^{n\alpha}|^2 |\phi_{jb}^{n\beta}|^2 \\
			B_{ab} \equiv & \gamma\sum_{n,\alpha\neq\beta} \phi_{ia}^{n\alpha}\phi_{ia}^{n\beta}\phi_{jb}^{n\alpha}\phi_{jb}^{n\beta} \\
			C_{ab} \equiv & \gamma\sum_{n,\alpha} |\phi_{ia}^{n\alpha}|^2 |\phi_{jb}^{n\alpha}|^2
		\end{align}
	\end{subequations}
	In~\cref{Supp:eq:nnCoulombMatrix} $U_0$ and $J_H$ refer to the Coulomb interaction strengths associated with the O~2$p$ orbitals. Following Ref.~\cite{liu2018pseudospin} we choose $U_0=0.7U_{avg}$ and $J_H=0.3U_0$ for honeycomb materials with octahedral oxygen structures. $A_{ab}, B_{ab},$ and $C_{ab}$, visible in~\cref{Supp:eq:WannierOverlap} are obtained from non-relativistic FPLO calculations. The Wannier function overlaps between a Ni $3d$ orbital $a$ on site $i$ and a O $2p$ orbital $\alpha$ on site $n$ is represented by $\phi_{ia}^{n\alpha}$. Finally, we introduced a scaling factor $\gamma$ to the overlap integrals, in order to account for screening of the Coulomb interactions with respect to the bare Slater integral. It acts as the fitting parameter to our model. The values obtained for the effective interaction matrices on the $Z$-bond in terms of $d_{xy}$, $d_{yz}$, $d_{xz}$, $d_{z^2}$ and $d_{x^2-y^2}$ orbitals are
	\begin{align}
		A = \begin{pmatrix}
			1.13 & 0.24 & 0.24 & 0.91 & 2.81 \\
			0.24 & 0    & 0.02 & 0.21 & 0.66 \\
			0.24 & 0.02 & 0 & 0.21    & 0.66 \\
			0.91 & 0.21 & 0.21 & 0.72 & 2.22 \\
			2.81 & 0.66 & 0.66 & 2.22 & 6.81
		\end{pmatrix}\text{meV}
	\end{align}
	\begin{align}
		B = -\begin{pmatrix}
			 0.71 &  0.01 &  0.01 & -0.90 & -2.73 \\
			 0.01 &  0    & -0.01 &  0.04 &  0.12 \\
			 0.01 & -0.01 &  0    &  0.04 &  0.12 \\
			-0.90 &  0.04 &  0.04 &  0.51 &  1.54 \\
			-2.73 &  0.12 &  0.12 &  1.54 &  4.70
		\end{pmatrix}\text{meV}
	\end{align}
	\begin{align}
		C = \begin{pmatrix}
			0.72 & 0.07 & 0.07 & 0.61 & 1.81 \\
			0.07 & 0    & 0.08 & 0.05 & 0.13 \\
			0.07 & 0.08 & 0    & 0.05 & 0.13 \\
			0.61 & 0.05 & 0.05 & 0.54 & 1.58 \\
			1.81 & 0.13 & 0.13 & 1.58 & 4.73
		\end{pmatrix}\text{meV}
	\end{align}
	\begin{align}
		\tilde{J} = \begin{pmatrix}
			 2.14 & 0.55 & 0.55 & 5.83 & 17.57 \\
			 0.55 & 0.01 & 0.36 & 0.31 &  0.94 \\
			 0.55 & 0.36 & 0.01 & 0.31 &  0.94 \\
			 5.83 & 0.31 & 0.31 & 1.52 &  4.50 \\
			17.57 & 0.94 & 0.94 & 4.50 & 13.40 \\
		\end{pmatrix}\text{meV}
	\end{align}

	\section{Supplementary Note 2: Extended Magnetic Couplings Data}
	
	\vspace{0.2cm}
	
	The parameter $\gamma$, which acts as screening for the 4\textsuperscript{th} order superexchange interaction, represents the only fitting parameter we introduced to our microscopic model of Na$_3$Ni$_2$BiO$_6$. In order to quantify its influence on the couplings, we present the 1\textsuperscript{st} NN magnetic couplings for several values of $\gamma$. The case $\gamma=0.28$ (see main text) is compared to fully turned off superexchange ($\gamma=0$) as well as to $\gamma=0.27$, which exhibits the 1/3-plateau phase at the experimentally observed field of $6.6$T, as visible in~\cref{fig:MagnetizationCurve_Gamma}~\textbf{a}.
    
	The data is presented in~\cref{Supp:tab:MagCouplingsExtended}.
	\begin{table}[h]
		\begin{tabular}{c|cc|cc|cc}
			& \multicolumn{2}{c|}{$\gamma=0$} & \multicolumn{2}{c|}{$\gamma=0.27$} & \multicolumn{2}{c}{$\gamma=0.28$} \\
			& $\eta=X$	& $\eta=Z$	& $\eta=X$	& $\eta=Z$	& $\eta=X$	& $\eta=Z$	\\
			\hline
			$J_1^\eta$		& -0.461	& -0.469	&  -3.529	& -3.657	& -3.643	&  -3.775	\\
			$K_1^\eta$		& -0.012	& -0.011	&  -0.017	& -0.015	& -0.018	& -0.016	\\
			$\Gamma_1^\eta$	& -0.004	& -0.004	&  -0.004	& -0.004	& -0.004	& -0.004 	\\
			$\Gamma_1^\eta$	&  0.003	&  0.003 	&   0.003	&  0.003	&  0.003	&  0.003	\\
			$B_1^\eta$		&  0.014 	&  0.011 	&   0.016	&  0.012	&  0.016	&  0.012	\\
		\end{tabular}
		\caption{1\textsuperscript{st} NN magnetic couplings varying with the screening parameter $\gamma$. Heisenberg $J$ is ferromagnetic for all values of $\gamma$ but increases significantly $\left(\Delta J_1^Z=-3.107\text{meV}\right)$ from $\gamma=0$ to $0.27$. All other couplings are only weakly influenced by the screening and stay within their respective order of magnitude.}
		\label{Supp:tab:MagCouplingsExtended}
	\end{table}

	\section{Supplementary Note 3: Extended Static Structure Factor Data}
	
	\vspace{0.2cm}
	
\begin{figure*}[t]
		\centering
		\includegraphics[scale=0.82]{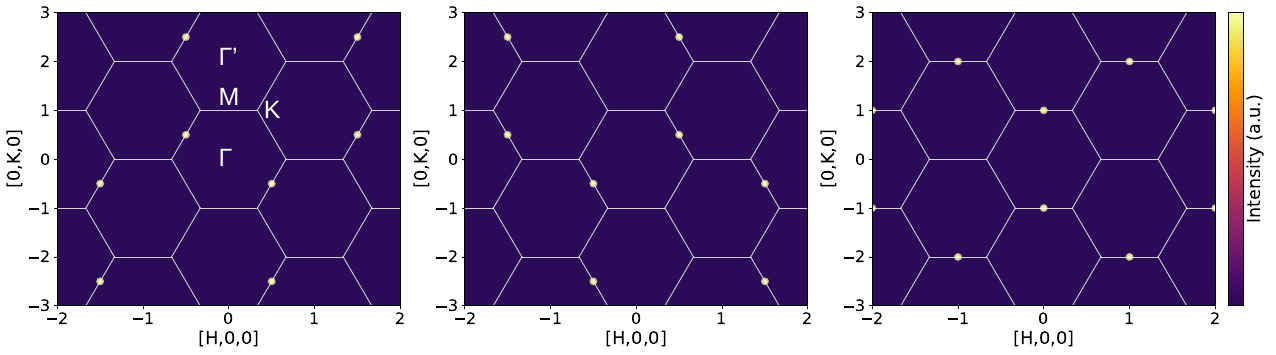}
		\caption{Static structure factor $S(\textbf{q})$ in zero-field environment, corresponding to the three possible ground state spin configurations (see~\cref{fig:SpinConfiguration}~\textbf{a}-\textbf{c}, in that order). Data is presented for the $(H,0,0)$, $(0,K,0)$ momentum space plane (reciprocal lattice directions of the conventional unit cell). The intensities are given in arbitrary units.}
		\label{Supp:fig:ENS_NoField}
	\end{figure*}

\begin{figure*}[t]
		\centering
		\includegraphics[scale=0.82]{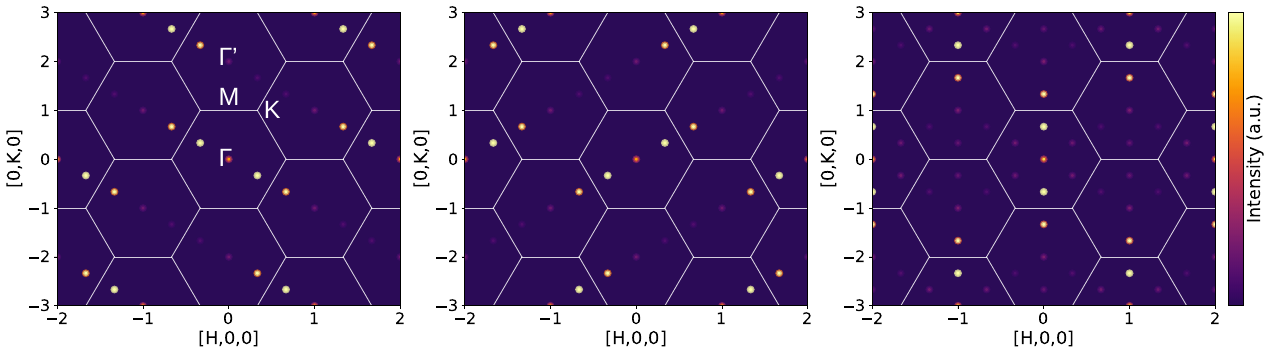}
		\caption{Static structure factor $S(\textbf{q})$ associated with the magnetic field $\mu_0H=5.5$T, $H\vert\vert c^*$ for the three possible ground state spin configurations (see~\cref{fig:SpinConfiguration}~\textbf{d}-\textbf{f}, in that order). Data is presented for the $(H,0,0)$, $(0,K,0)$ momentum space plane (reciprocal lattice directions of the conventional unit cell). The intensities are given in arbitrary units.}
		\label{Supp:fig:ENS_Field}
	\end{figure*}

	The SSF results, presented in the main part of the text, each consist of the sum of three separate data sets, one for each of the three possible spin configurations.~\cref{Supp:fig:ENS_NoField} and~\cref{Supp:fig:ENS_Field} show the spectra of the individual spin configurations.

	\clearpage

\end{document}